\documentclass[apj]{emulateapj}

\usepackage{amsbsy}

\usepackage{amsmath}
\usepackage{amsthm}
\usepackage{eucal}
\usepackage{amssymb}
\usepackage{mathrsfs}

\usepackage{color}

\graphicspath{{/Users/patou/Documents/RAPPORT_BOULOT/NANO-POUSSIERES/FIGURES/Figures_article_dust_heliosphere/}}

\shorttitle{Nanodust through the heliosphere measured by Cassini}
\shortauthors{Schippers et al.}

\begin{document}

%
%
%
\setkeys{Gin}{draft=false}
%

%
%


\title{Nanodust detection between 1 and 5 AU by using Cassini wave measurements}


%
%


%



\author{P. Schippers\altaffilmark{1}, N. Meyer-Vernet\altaffilmark{1}, A. Lecacheux\altaffilmark{1}, S. Belheouane\altaffilmark{1}, M. Moncuquet\altaffilmark{1}, \\ W.S. Kurth\altaffilmark{2}, I. Mann\altaffilmark{3}, D. G. Mitchell\altaffilmark{4}, and N. Andr\'e\altaffilmark{5}}
%
%
\altaffiltext{1}
{LESIA - CNRS - Observatoire de Paris, 5 place Jules Janssen, 92195 Meudon, France.}
\altaffiltext{2}
{Department of Physics and Astronomy, University of Iowa, Iowa City, Iowa, USA.}
\altaffiltext{3}
{EISCAT Scientific Association, Kiruna, Sweden and Department of Physics Ume\aa{} University, Sweden.}
\altaffiltext{4}
{Applied Physics Laboratory, John Hopkins University, Laurel, Maryland, USA.}
\altaffiltext{5}
{IRAP, 9 Avenue du Colonel Roche, 31028 Toulouse, France.}

%

%
%

\begin{abstract}

The solar system contains solids of all sizes, ranging from km-size bodies to nano-sized particles. 
Nanograins have been detected \textit{in situ} in the Earth's atmosphere, near cometary and giant planet environments, and more recently in the solar wind at 1 AU. 
These latter nano grains are thought to be formed in the inner solar system dust cloud, mainly through collisional break-up of larger grains and are then picked-up and accelerated by the magnetized solar wind because of their large charge-to-mass ratio.
In the present paper, we analyze the low frequency bursty noise identified in the Cassini radio and plasma wave data during the spacecraft cruise phase inside Jupiter's orbit. The magnitude, spectral shape and waveform of this broadband noise is consistent with the signature of nano particles  impinging  at nearby the solar wind speed on the spacecraft surface. 
Nanoparticles were observed whenever the radio instrument was turned on and able to detect them, at different heliocentric distances between Earth and Jupiter, suggesting their ubiquitous presence in the heliosphere. We analyzed the radial dependence of the nano dust flux with heliospheric distance and found that it is consistent with the dynamics of nano dust originating from the inner heliosphere and picked-up by the solar wind. The contribution of the nano dust produced in asteroid belt appears to be negligible compared to the trapping region in the inner heliosphere. In contrast, further out, nano dust are mainly produced by the volcanism of active moons such as Io and Enceladus.

\end{abstract}


\section{Introduction}

 The interplanetary space contains dust particles over a large range of sizes. They are released through the activity of comets and moons and fragmentation and impacts of asteroids. Until recently the smaller particles detected in the  interplanetary dust cloud were several tens nanometers across. Smaller nanograins have been remotely identified in the interstellar medium a long time ago through different mechanisms such as far UV extinction, optical luminescence, and infrared emission. Such remote detection is however not as effective for solar system nanodust mainly because of the weak optical depth \citep{Li2012}, so that in situ detection provides a better opportunity. Such grains have peculiar properties due to their very small size. In particular, their large surface to-volume ratio favors the surface exchange reactions with their environment, and their large charge-to-mass ratio allows them to be easily accelerated in large-scale electric fields that occur in planetary magnetospheres \citep{Burns2001} and/or close to the moons \citep{Farrell2012}. But they can also be accelerated by the magnetized solar wind, though on longer time scales \citep{Mann2007} by a process that is akin to the pick-up of freshly produced ions in the solar wind \citep{Luhmann2003}. These nano dust particles are then expelled from the inner heliosphere to large distances \citep{Hamilton1996,Czechowski2010}.
 
Nano grains have been so far detected \textit{in situ} in cometary environments \citep{Utterback1990}, in the solar wind at 1 AU by \cite{MeyerVernet2009b,Schippers2014}, and in the environments of giant planets such at Jupiter and Saturn \citep[and references therein]{Hsu2012}. The presence of a reservoir of nanograins at 0.2 AU has been theoretically predicted by \cite{Mann2007} and is suspected to be the main source of nanodust in the inner heliosphere.
{Indeed, dust is generated by mutual collisions in the solar system dust cloud and production rate is expected to increase towards the Sun where increasing number density and relative velocities lead to a maximum in the production of the dust fragments.  In the very close vicinity of the Sun, the {interplay of gravity and electromagnetic forces} leads to a trapping of the dust and this occurs under certain conditions at approximately 0.2 AU \citep{Czechowski2010}}.
In the outer heliosphere, the nano dust sources so-far identified are the giant planet moons Io \citep{Maravilla1995} and Enceladus \citep{Spahn2006} which are geologically active and release dust through volcanism and cryovolcanism respectively.  

Besides dust analyzers which are designed to measure \textit{in situ} the micron sized grains \citep{Auer2001}, radio wave instruments have the ability to detect dust grains \citep{Oberc1996}. This technique is used in the present paper. It is based on the principle that the high-speed impacts of grains on the spacecraft induce electric pulses that are recorded by the wave receivers \citep{MeyerVernet2015}. 
\cite{MeyerVernet2009b} reported the first nano dust detection at 1 AU with the radio instrument on STEREO (Solar Terrestrial Relations Observatory). 
This detection was made possible by the high-speed of these particles which move at nearly the solar wind speed due to their high charge-to-mass ratio \citep{Mann2014}.
More recently, \cite{Schippers2014} confirmed this discovery using the radio measurements of the Cassini mission when it flew close to Earth orbit in August 1999.
The radio and plasma wave instrument (RPWS) onboard Cassini was episodically turned on during its cruise phase and recorded signatures similar to those observed by this instrument  when Cassini encountered high-speed nano dust near Jupiter \citep{MeyerVernet2009a}.
   
The aim of the present paper is to analyze these signatures and understand the origin of the dust. 
In section 2 we recall the method of nanodust detection with radio wave instrumentation. In section 3, we apply the method to identify the dust and estimate the flux with radio wave data onboard Cassini. The results are discussed in section 4.

\section{In situ Dust detection with a radio receiver}

\subsection{Basics of dust detection with wave measurements}
Radio and plasma wave instruments have been traditionally used to measure electromagnetic radiations and also to monitor electron bulk properties such as density and temperature through the analysis of electrostatic fluctuations induced by the charged particle thermal motion around the antennas (i.e. thermal-noise spectroscopy) \citep{MeyerVernet1979}.  
Although radio and plasma wave receivers are not designed for dust measurements, these have the capability of revealing the presence of micro and nano dust (within some conditions).
Indeed, electric antennas are sensitive to dust because the high speed impact of a dust grain on the spacecraft surface generates a plasma cloud whose expansion creates charge decoupling \citep{Drapatz1974} . This method is complementary to conventional impact ionization dust detectors, which provide more information per impact but are limited to the small size of the detectors. 
In the solar wind, the spacecraft is positively charged because of photoelectron emission so that  the electrons from the expanding cloud are recollected by the target whereas the expanding ions induce a potential on the spacecraft and antennas \citep{MeyerVernet2014}.
This holds until the density of the cloud reaches the density of the ambient plasma, so that the characteristic time of this process satisfies:
\begin{equation}
 \tau_r \lesssim (3Q/(4 \pi en))^{1/3}/v_E
 \label{rise_time}
 \end{equation}
 where $Q$ is the charge of the plasma cloud, $n$ is the ambient density and $v_E$ is the expansion velocity \citep{MeyerVernet2009a,MeyerVernet2009b}.
The electric signal measured by the antennas can then be used as a diagnosis of the dust. This technique is very sensitive because of the large detection area and solid angle since the whole spacecraft surface is the target.     
It was first developed to analyze micro-dust at Saturn using the radio instrument onboard Voyager \citep{Aubier1983, Gurnett1983, MeyerVernet1996}, and extended to fast nano dust detection at Jupiter \citep{MeyerVernet2009a} using the wave data from Cassini, and at 1 AU using STEREO/WAVES \citep{MeyerVernet2009b} and Cassini/RPWS \citep{Schippers2014} .   

The maximum voltage amplitude between one antenna and the spacecraft produced by the impact of one dust grain on the spacecraft surface writes: 
\begin{equation}
\delta V\simeq \Gamma~Q/C 
\label{Eqn_Single_Pulse}
\end{equation}
where $Q$ is the charge generated by the dust impact, $C$ is the capacitance of the spacecraft surface and $\Gamma$ the antenna gain ($\simeq$ 0.4 in this case). 
The impact charge $Q$ depends on the grain mass and speed with various relationships of the form $Q\propto$m$^{\alpha} v^{\beta}$ with $\alpha \simeq 1$ and $\beta \simeq 3 - 4.5$. The coefficients depend on mass, speed, angle of incidence, as well as grain and target composition  \citep{Goeller1989,Burchell1999} and have not been measured for either nano dust or $v>$70 km/s \citep{Auer2001}.

According to \cite{McBride1999}, the charge released can be approximated by 
\begin{equation}
Q=  0.7mv^{3.5}
\label{McBride}
\end{equation}
where $m$  and $v$ are respectively the mass and the speed of the grain. 
This method is then sensitive to massive dust particles and/or fast particles (speed higher than a few km/s) as the induced charge is strongly dependent on the speed. 
It is however not able to determine separately the speed and the mass of the particles. 
Note that impact ionization yields of materials relevant for Cassini have been reported for micro dust at speeds $<$40 km/s \citep{Collette2014}. Extrapolating these results for nano dust impacting at several hundreds of km/s produces charges of the same order of magnitude as Eq. (\ref{McBride}).

The power spectrum of pulses of amplitude (\ref{Eqn_Single_Pulse}) with impact rate unity is: 
\begin{equation}
V_{fi}^2\simeq 2(\Gamma Q/C)^2 \omega^{-2} (1+\omega^2 \tau_r^2)^{-1}
\label{Eqn_Vfi}
\end{equation}
where $\tau_r$ is the pulse rise time and $\tau_d$ $>> 1/\omega$ the pulse decay time \citep{MeyerVernet1985}.
When the rise time exceeds $1/\omega$, which occurs in low density plasma such as solar wind, the signal is then proportional to 
\begin{equation}
V_{fi}^2\propto \omega^{-4}
\label{Eqn_f4} 
\end{equation}
For a cumulative distribution of the grain flux $F(m)$ over a surface $S$, the expected power spectrum is 
\newline

\begin{equation}
V_f^2\simeq S \int_{m_{min}}^{m_{max}} dm \frac{dF(m)}{dm} V_{fi}^2(m) 
 \label{Eqn_Total_Dust}
\end{equation}

\subsection{Plasma Wave Instrument onboard Cassini}

The radio and plasma wave instrument onboard Cassini \citep{Gurnett2004} is composed of five detectors connected to three 10m length electric antennas. The present analysis is mainly  based on the High Frequency Receiver (HFR) which acquires the power spectral density from 3.7 kHz to 16 MHz (in the present study we only use the lowest filter ranging up to 16 kHz). We also analyze time series measured by the Five-Channel Waveform receiver which acquires waveforms in passbands of either 1 to 26 Hz or 3 Hz to 2.5 kHz.     
On Cassini, the antennas are positioned symmetrically with respect to the spacecraft body, which means that we expect that measurement in "monopole" mode (when the potential is measured between one antenna and the spacecraft body) should be similar for all the antennas \citep{MeyerVernet2014}. As a consequence, the voltage measured in "dipole" mode (potential difference between two antennas) should be negligible in the absence of other sources of electric field and of dust impact on one of the antennas. 
On Cassini, one of the antenna is always connected in monopole (w), while the two others antennas are either connected in monopole or dipole (u-v).     
In the present study, we will thus use the absence of signal in dipole mode to ensure that the data used are not contaminated by plasma and electromagnetic waves and base the dust measurements on data acquired in monopole mode.

\section{Dust identification and flux calculation}

\subsection{Wave Dataset}

During its phase cruise to Saturn, the instruments onboard Cassini were turned on several times discussed in turn below.  

\subsubsection{The ICO period: 1.6 AU}

From December 1998 to January 1999, the instruments underwent a commisioning phase (``ICO'').  

 {Figure \ref{Figure_Spectra}a} displays the electric power density spectrogram (PDS) in $V^2/m^2/Hz$ acquired by the HFR receiver in monopole mode (w) on January 4 of 1999 between 00:00UT and 18:00UT, when the spacecraft was at a heliospheric distance of 1.56 AU. 
We observe 1) a continuous background noise at low frequency, due to the impact of the plasma on the spacecraft, and 2) a sporadic and bursty broadband signal. 
{Figure \ref{Figure_Spectra}b} shows the voltage power spectral density at two different times on day 1999-004: inside and outside a ``burst'' event. At 09:54:30 UT (outside), the signal is well modeled by a power law $\propto \omega^{-2}$, typical of plasma (``shot'') noise signature in both amplitude and spectral index \citep{MeyerVernet1989}.  At 10:07 UT,  the voltage power is enhanced by two orders of magnitude, and the slope is steeper than the plasma background noise. We have superimposed two approximate models of dust impact power spectrum with Equation \ref{Eqn_Vfi} and Equation \ref{Eqn_f4}, respectively in red and green.

Figure \ref{Figure_Waveform} displays the electric field time series acquired by WFR in monopole mode the same day, at 10:06:34 UT.
We observe that:
\begin{enumerate}
\item The signal displays a series of spikes typical of dust impact ionization. The complicated shape is expected to be produced by the electric field induced by the impact ions, the recollected electrons, and the finite charging time scale of the spacecraft/antenna/receiver system.
\item The amplitude of the electric field in the spike $\delta E=\delta V/L \simeq$ 1 mV/m is consistent with the potential (Eq. \ref{Eqn_Single_Pulse}) produced by the impact of particles of a few nanometers at about the solar wind speed, (with L=10m, C=200pF, and the induced charge given by Eq. \ref{McBride}).
\item The occurrence of the spikes is consistent with the flux on the spacecraft surface $S\simeq 15$m$^2$ of interplanetary nano grain predicted by extrapolating the \cite{Grun1993} model to the sub-micron size range using the collisional fragmentation law $F(m)\propto m^{-5/6}$.
\end{enumerate}  

\subsubsection{Near Earth's orbit: 1 AU}

From August 15 and September 15 of 1999, the Cassini instruments were turned on during the Earth flyby (19 September 1999). \cite{Schippers2014} reported the first nano dust detection at about 1 AU with Cassini using RPWS/HFR data during this period, whose results confirmed the STEREO observations by \cite{MeyerVernet2009b}. 
  
\subsubsection{In the asteroid belt: 2.9 AU} 

Between February 23 and March 3 of year 2000, the radio instrument was turned on again. Cassini spacecraft was then around 2.9 AU, within the Asteroid belt.
{Figures \ref{Figure_Spectra}c and d} display the same information as in {Figures \ref{Figure_Spectra}a and b} for day 27 of February 2000. In this data sample, we observe that 
 \begin{enumerate}
 \item the plasma noise level (proportional to the plasma density) is smaller than in Figure \ref{Figure_Spectra}b by a factor of about 3, equal to the the solar wind density decrease between 1.6 and 2.9 AU, 
 \item the dust noise intensity is extremely variable, 
 \item the dust noise is closer to the $\omega^{-4}$ power law model. This last point is consistent with the first since the rise time (Equation \ref{rise_time}) is expected to increase as the local plasma density decreases.
\end{enumerate}

\subsubsection{Beyond the asteroid belt up to Jupiter's orbit}

Before Cassini reached its closer position to Jupiter (30 december 2000),  the instrument was turned on for about four months. During this period the RPWS/HFR mode was however not adequate to detect nanograin impacts because at closer distance to the planets, the instrument configuration was adapted in order to measure planetary radioemissions at a high rate. For doing so, the integration time $\Delta t$ was set to a value four times smaller than during the previous period. Since the receiver averages the spectral density over the integration time, grain detection requires that the impact rate be high enough to ensure at least one impact during this time.

 Unfortunately, $\Delta t$ then becomes so small that during this period, virtually no nanograin can impact the spacecraft surface except if the flux is much greater than the model (black line in Figure 5). Indeed, with a cumulative flux $\propto m^{-5/6}$, the maximum mass impacting during $\Delta t$, given by $F(m_{max})S \Delta t \simeq 1$, is smaller than that of 1 nm-grains. A possible explanation for the lack of subnanometer grains is that the electric field at the grain surface becomes so high that the electrostatic stress can exceed the maximum grain cohesion strength, causing the grain's explosion \citep{MeyerVernet2015}.

\subsection {The high energy particle detector discharge issue}
During the early mission, the Magnetospheric Imaging Instrument (MIMI) \citep{Krimigis2004} onboard Cassini reported an issue: discharges of the Imaging 
Neutral Camera (INCA), possibly triggered by dust impacts and/or sunlight shining on the charged particle rejection plates \citep{Schippers2014}.
These discharges were thought to be produced by the negative high voltage applied to the plates. Evidence was found that RPWS data were contaminated by MIMI/INCA discharges \citep{Schippers2014}. 
 A discharge monitor was incorporated in the MIMI instrument to identify and discard the corrupted data. It takes into account a couple of noise indicators identifiable in a few control parameters of INCA. Using this device led us to remove short time intervals possibly affected by contamination by these discharges. 
This type of contamination was strongly reduced after the MIMI instrument team decided to turn off the negative voltage mode on day 13 of 2001. 

\subsection{Nanodust Flux determination}

To calculate the flux of nano particles from the measured power spectral density from Equation (\ref{Eqn_Total_Dust}), we need:
\begin{enumerate}
\item To assume a power index for the distribution function $F(m)$. Here we choose the low mass extrapolation of the interplanetary dust model by \cite{Grun1985}, which yields the cumulative distribution
\begin{equation}
 F(m)=F_0m^{-5/6} 
 \label{Eqn_Cum_flux}
 \end{equation}
In that case, Equations (\ref{Eqn_Vfi}) and (\ref{Eqn_Total_Dust}) lead to:
 \begin{equation}
 {V_f^2\simeq 0.7 S (\Gamma/C \tau_r)^2 v^7 F_0 m_{max}^{7/6}/\omega^4}
 \label{Eqn_V_F0}
 \end{equation}
 where $S$ is the spacecraft surface ($\simeq$ 15 m$^2$), $m_{min}$ and $m_{max}$ are the minimum and maximum detected {mass in kg, $v$ is the speed in km/s}, and the approximation (\ref{Eqn_f4}) is valid for $\omega \tau_r>>1$.  
 
 \item To consider the speed of the grain particles  $\mathbf{v_d}$. Numerical simulations show that farther than 1 AU, grains of radius smaller than about 10 nm are accelerated at a speed close to
 \begin{equation}
  \mathbf{v_d}= \frac{( \mathbf{v_{sw}}\times  \mathbf{B})\times  \mathbf{B}}{B^2}
 \label{Eqn_speed}
 \end{equation}
 where  $\mathbf{v_{sw}}$ is the solar wind speed and $\mathbf{B}$ is the interplanetary magnetic field \citep{Czechowski2012,MeyerVernet2015}.
 According to Equation (\ref{Eqn_speed}), we estimate a nano grain impact speed of 360 km/s at 1.6 AU and 450 km/s  at 2.9 AU.

\end{enumerate}

At each time, we calculate the flux $F_0$ by equating the expression (8) to the measured spectral power \citep{Schippers2014}.
In order to eliminate the contributions of discharges, radio and plasma emissions, and plasma quasi-thermal noise, we have carefully selected the times when the following conditions are met:
\newline 1) MIMI/INCA indicators do not show evidence of the presence of discharges (from Section 3.2),
\newline  2) the average index of the power law adjusted to the measured spectrum lies between $-$2.2 and $-$4, as expected for dust impacts (from Section 2.1), 
\newline 3) the level of the three monopoles are similar {(within 50$\%$)} when they are available \citep{MeyerVernet2009a} as expected for pulses in spacecraft potential, and
\newline  (4) the ratio between the measurements in monopole and dipole mode is greater  than a factor of 3. This condition enables us to eliminate contributions from plasma wave electric fields.

The flux calculation and criteria were applied on the data acquired during the Cassini ICO period (January 4 of 1999) and the asteroid belt crossing (Feb 23 of 2000 - Mar 3 of 2000). 
Figure \ref{Figure_F0_time}a displays the power spectral density for asteroid belt crossing. The corresponding normalized flux $F_0$ is plotted in Figure \ref{Figure_F0_time}b with black crosses with the discharge intervals indicated in grey vertical lines. The horizontal line represents the flux $F_0$ averaged on the whole time interval. Figure \ref{Figure_F0_time}c displays the histogram of the flux values.  It shows a gaussian distribution around F$_0\simeq 10^{-18.75\pm 0.23}m^{-2}s^{-1}kg^{5/6}$, with a slight excess towards large values.  Note the lack of data points in the low flux part of the distribution which is expected to be due to the threshold effect of the criteria described above and to the small integration time $\Delta$t. 
 The average normalized flux calculated from data acquired at 1.6 AU is  F$_0\simeq 10^{-18.12\pm 0.11} m^{-2}s^{-1}kg^{5/6}$, whereas the average normalized flux determined at 1 AU \citep{Schippers2014} is F$_0\simeq 10^{-17.94\pm 0.14} m^{-2}s^{-1}kg^{5/6}$.

 \section{Radial distribution of nano dust in the heliosphere and Discussion}
 
 Figure \ref{ComparisonFlux} displays the intervals when the instrument was on and able to detect  nano dust  because the integration time was large enough (marked in green) and when it exhibits nano dust signatures (marked in red).
One sees that nano dust is observed whenever the instrument is capable of detecting them.  

 Our determination of the nano dust flux at three Cassini positions (1, 1.6 and 2.9 AU) in the solar wind with Cassini gives for the first time an estimate of the nano flux radial distribution 
 and hints about the dynamics in the inner heliosphere.
In Figure \ref{ComparisonFlux}, we display the cumulative nano dust flux deduced from the data for particles of mass 10$^{-20}$kg at the three positions (in red); the solid vertical extension of the bars corresponds to the mean flux value $\pm 1 \sigma$ and the dotted bars range from the mean flux $\pm 3 \sigma$.  
The superimposed black solid line represents a r$^{-2}$ power law model starting from the cumulative flux at 1 AU, F$_{1AU}$, determined by \cite{Schippers2014} at m=10$^{-20}kg$.  
The other symbols denote measurements close to Jupiter. 

The crosses, triangles and diamonds show data from the Cosmic dust analyzer aboard Ulysses and Galileo \citep{Kruger2006}, whereas the square stems from Cassini/RPWS \citep{MeyerVernet2009a}. The color code refers to the distance of the spacecraft with respect to Jupiter. 
  These measurements around 5 AU were attributed to high-velocity dust streams ejected by Jupiter  \citep{Grun1993, Grun1996,Kruger2006}.
Various observations by dust detectors onboard Ulysses, Galileo and Cassini during spacecraft encounters with Jupiter together with trajectory simulations \citep{Zook1996,Graps2001} have shown a grain size of about 5-10 nm (i.e. mass $\simeq$ 10$^{-21}$-10$^{-20}$kg) with a speed of 200-450 km s$^{-1}$.

Our result reveals that:
\begin{itemize}

\item The nano dust appear ubiquitous in the interplanetary medium at least near the ecliptic between 1 and 5 AU.

\item  The interplanetary nano dust flux decreases outwards suggesting production in the innermost heliosphere, inside Earth's orbit, different from the timely and spatially limited sources such as comet and planets. This result is consistent with \cite{Mann2007} who suggested the existence of a trapping/source region for the nano dust very close to the Sun (0.2 AU), picked-up and accelerated by the magnetized solar wind \citep{Czechowski2010,Czechowski2012,Mann2012}. 

\item The observations confirm that the interplanetary nano dust flux follows a power law trend close to r$^{-2}$ outwards of 1 AU. 
This variation was expected since the mean velocity of the particles is expected to change relatively weakly after they have been accelerated close to the solar wind drift velocity, so that the flux conservation yields a decrease in proportion of the inverse squared heliocentric distance.
This result is consistent with numerical simulations \citep{ThSoraya} of nano dust dynamics. 
Figure \ref{Figure_simu} shows the result of numerical simulations of the nano dust dynamics for a source located in the inner heliosphere at 0.2 AU \citep{Mann2012}. The dashed and dotted lines correspond to an inclination of the interplanetary current sheet of 70$^{\circ}$ (as observed in 1999-2001) for two different polarity orientations of the magnetic field: the dashed line corresponds to a focusing electric field $\mathbf{v} \times \mathbf{B}$ (i.e. pointing to the heliospheric current sheet) and the dotted line corresponds to a defocusing electric field (i.e. pointing away from the current sheet). 
We observe that the simulated flux radial dependence is very close to the $r^{-2}$ power law model displayed with a black solid line.  

\item The inner heliospheric source produces the main contribution to the nano dust flux within 5 AU, except very close to Jupiter. The asteroid belt contribution appears to be negligible. Note in particular that Cassini was located more than 100$^{\circ}$ away from the dwarf planet Ceres \citep{Kuppers2014} so that we were not able to quantify its production/contribution.   
  
\item Closer than 2 AU from Jupiter, the flux is much higher, corresponding to the nanodust streams of Jovian origin \citep{Zook1996}, the dominant source of nano dust in the planet's vicinity \citep{Graps2000}. 
 
\end{itemize}

\section{Conclusion} 

We analyzed the radio and wave data onboard Cassini from 1997 to 2001 (first part of its cruise phase) to identify nano dust signatures in the interplanetary medium. 
 {While the instrument and the operating modes were not defined for this purpose,} evidence for nano dust was observed whenever the instrument was turned on with a time integration large enough to enable dust detection. This took place during a few weeks around at  three different heliospheric distances: 1 AU, 1.6 AU and 2.9 AU. The observed flux distribution is consistent with nano dust produced in the inner heliosphere as suggested by \cite{Mann2007}, picked-up by the magnetized solar wind and filling the heliosphere \citep{Czechowski2010,Czechowski2012,Mann2012}.  
{This study presents the first results on nano dust detection on a large radial distance range in the interplanetary medium, obtained serendipitously by a single spacecraft during its phase cruise .
The spatial coverage of the results is however limited due to operating modes that were not adapted to the nano dust detection. Incidentally this shows the importance of the data acquired during mission cruise phases. }

%
%

\begin{acknowledgments}
The data are from the RPWS/HFR receiver of Cassini and are hosted in LESIA, Observatory of Paris. 
We acknowledge B. Cecconi for his assistance in providing the data.
The research at LESIA (Observatory of Paris) is supported by the CNES (Centre National d'Etudes Spatiales) agency. The research at the University of Iowa was supported by NASA through contract 1415150 with the Jet Propulsion Laboratory.
 
 
\end{acknowledgments}



 \setcounter{figure}{0}

\begin{figure}[htbp]
\includegraphics[width=42pc]{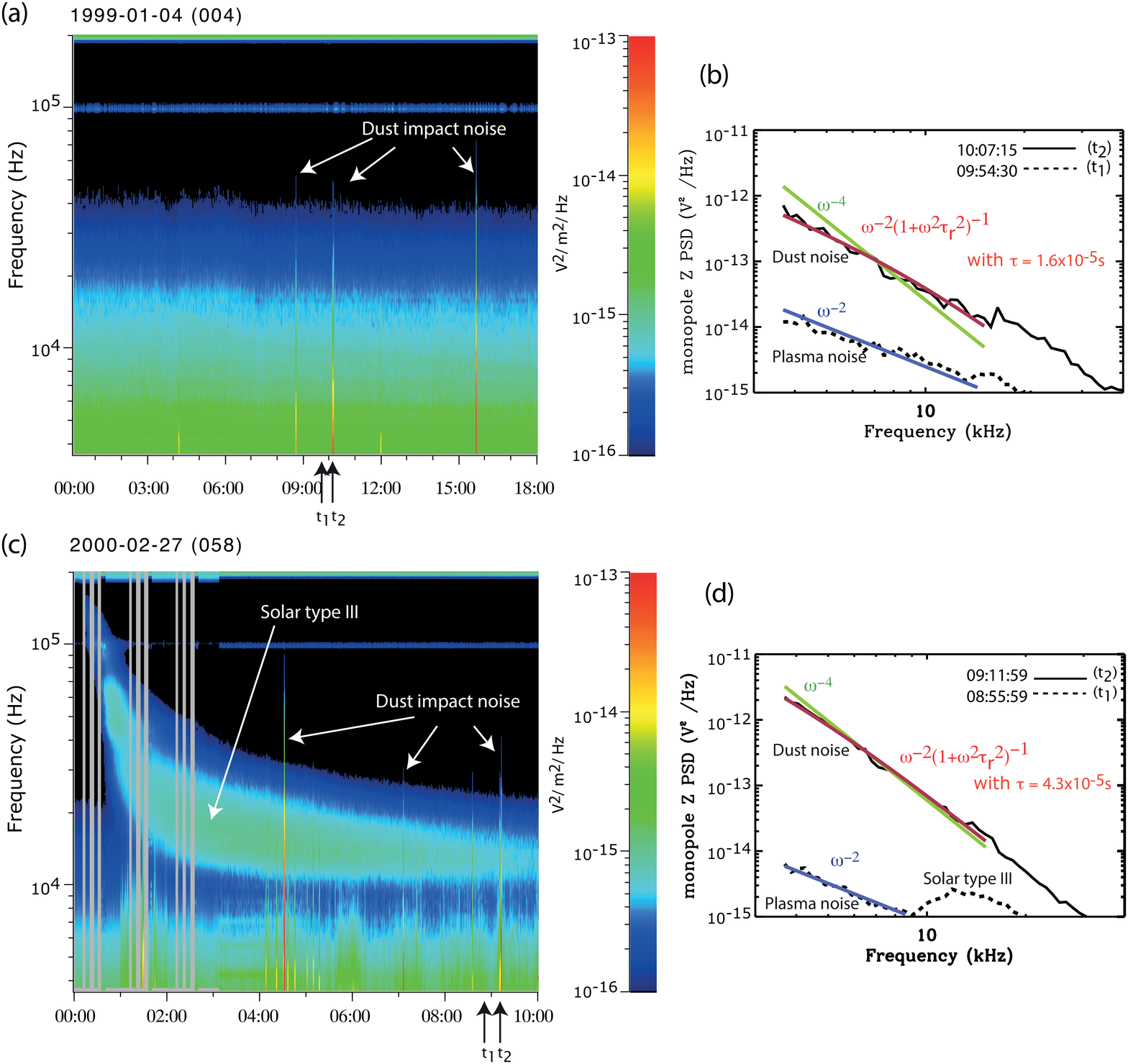}
\caption{Panel a: Electric Power spectral density as a function of frequency and time on day 004 of 1999 between 00:00 and 18:00 UT acquired with the antenna Z in monopole mode. 
Panel b: Voltage power spectral density spectra at 09:54UT and 10:07UT, in solid and dashed lines respectively. Three power spectral density models are superimposed: 1) the dust impact model: V$_f^2$ $\propto\omega^{-2}(1+\omega^2\tau_r^2)^{-1}$ (in red);  2) the dust impact model with $\tau_r >> 1/\omega$: V$_f^2$ $\propto\omega^{-4}$ (in green);  and 3) the plasma noise model: V$_f^2$ $\propto\omega^{-2}$ (in blue). Panel c and d: Same as Panel a and b on day 058 of 2000 between 00:00UT and 10:00UT.}
\label{Figure_Spectra}
\end{figure}

\begin{figure}[htbp]
\includegraphics[width=28pc]{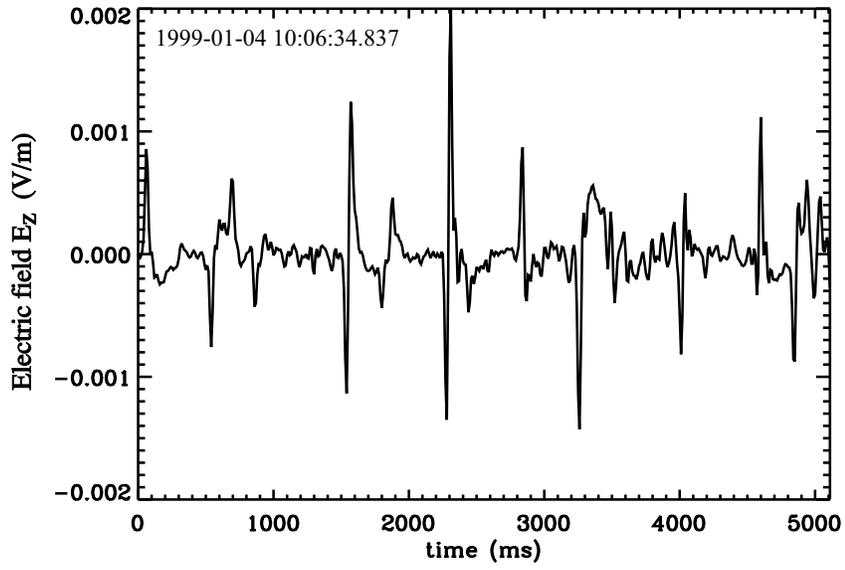}
\caption{Electric field time series acquired by the RPWS/Waveform receiver (WFR) on Jan 4 of 1999 on monopole w. The signal displays spiky signatures typical of nano dust impacts on the spacecraft \citep{Zaslavsky2012} .}
\label{Figure_Waveform}
\end{figure}

\begin{figure}[htbp]
\includegraphics[width=42pc]{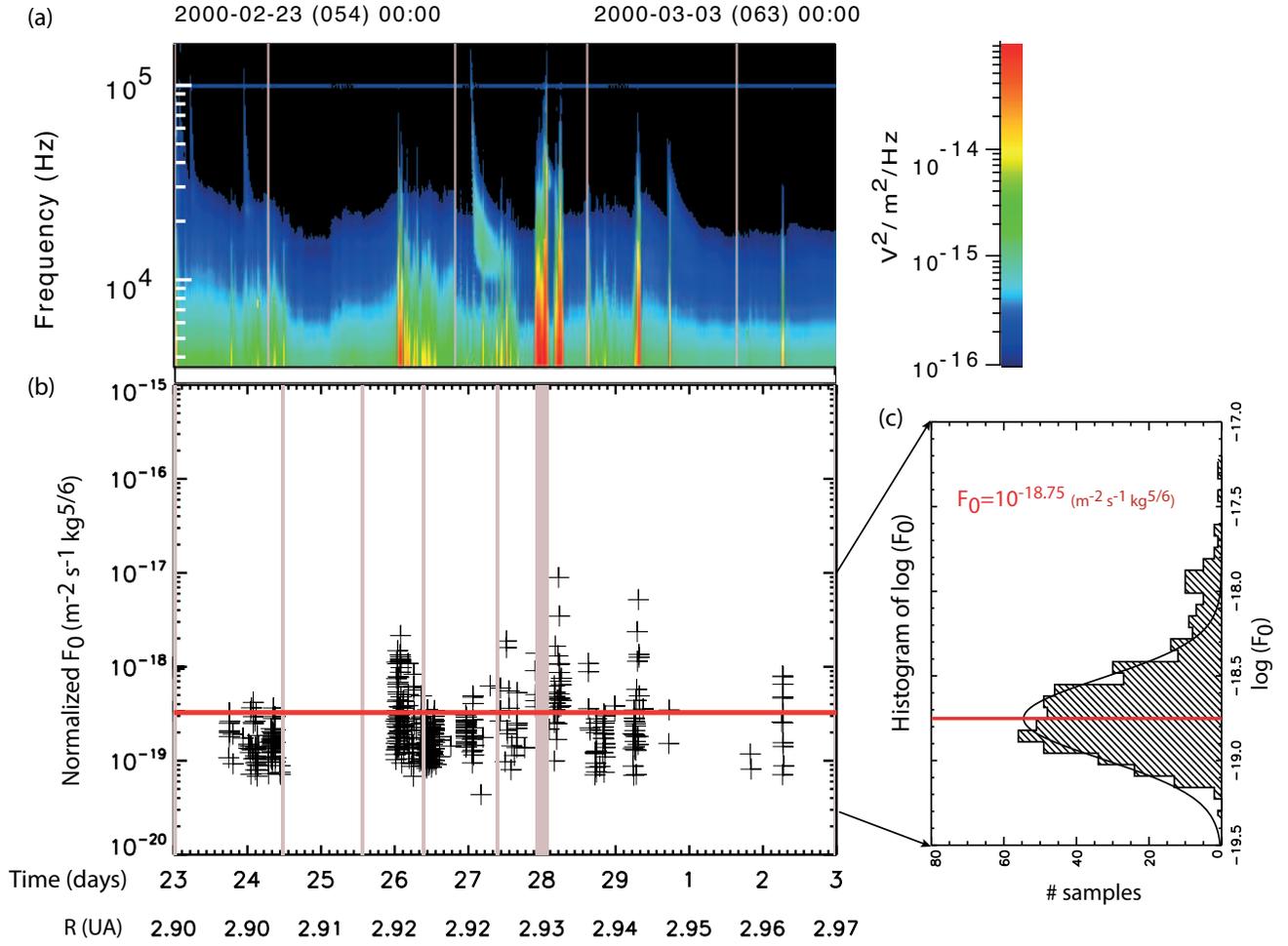}
\caption{Panel a: RPWS/HFR spectrogram of electric field below 200 kHz between Feb 23 and Mar 03 of 2000. Panel b: Normalized nanodust flux F$_0$ calculated by equating the theoretical power spectra model expected for a dust mass distribution flux F=$F_0m^{-5/6}$ from dust collisional fragmentation \citep{Dohnanyi1969} and the measured power V$_f^2$ at 10 kHz. Grey vertical lines indicate time intervals when MIMI/INCA discharges were identified (discarded from our dataset). Panel c: Histogram of F$_0$ for the full time interval.} 
\label{Figure_F0_time}
\end{figure}

\begin{figure}[htbp]
\includegraphics[width=.7\textwidth]{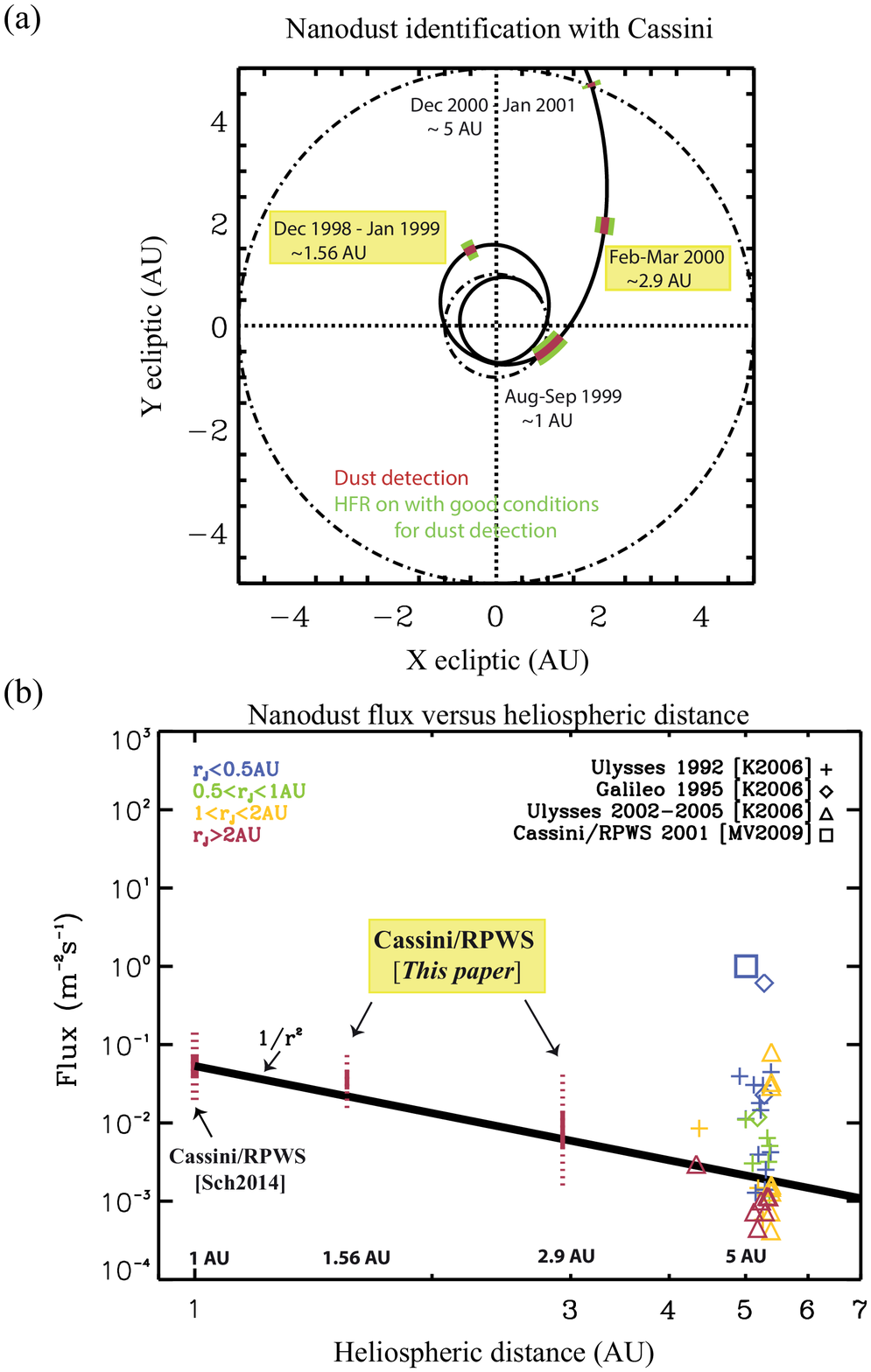}
\caption{Panel a: Cassini trajectory (black solid line) with periods when RPWS/HFR detector was able to detect dust (in green) and when we actually observe nanodust (in red). 
With Cassini, we were able to identify nanodust whenever the instrument could measure them: after Earth flyby at 1 AU \citep{Schippers2014} (confirmation of STEREO results by \cite{MeyerVernet2009b}, at 1.56 AU [this paper], at 2.9 AU [this paper] and 5 AU \citep{MeyerVernet2009a}. Panel b: Radial dependence of nanodust flux with mass=10$^{-20}$kg. Cassini measurements (except at 5 AU) appear to follow the decreasing trend $\propto \frac{1}{r^2}$ (solid black line) predicted by simulations (see Figure \ref{Figure_simu}). At 5 AU, we superimposed measurements from \cite{Kruger2006} obtained with Galileo and Ulysses. The references [K2006], [MV2009] and [Sch2014]  stand for \cite{Kruger2006}, \cite{MeyerVernet2009a} and \cite{Schippers2014} respectively.}
\label{ComparisonFlux}
\end{figure}

\begin{figure}[htbp]
\includegraphics[width=28pc]{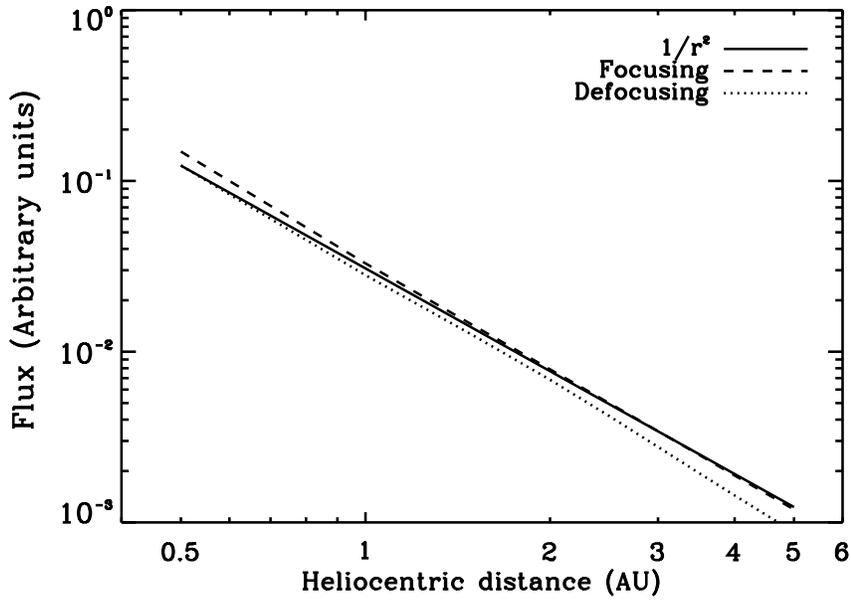}
\caption{Simulation of nanodust flux distribution across the heliosphere for a dust source located near 0.2 AU, in focusing and defocusing electric field configuration (respectively pointing towards and away from the current sheet). Dust flux radial dependence is very close to the power law $1/r^2$  (solid line).}
\label{Figure_simu}
\end{figure}


\end{document}